\documentclass[12pt]{article}
\usepackage{amsmath}
\usepackage{amssymb}

\ifx\pdftexversion\undefined
  \usepackage[dvips]{graphicx}
\else
  \usepackage[pdftex]{graphicx}
\fi

\advance\oddsidemargin by -0.5in 
\advance\textwidth by 1in
\def\sech{{\rm sech}}

\def\arctanh{{\rm arctanh}} 

\begin{document}

\begin{titlepage}
 \date{}
\title{Perturbation theory for localized solutions of sine-Gordon equation: decay of a breather and pinning by microresistor}

\author{D. R. Gulevich and F. V. Kusmartsev \\
 {\it Department of Physics, Loughborough University, Loughborough, LE11 3TU, UK}}
 \maketitle
 \begin{abstract}
 
We develop a perturbation theory that describes bound states of solitons localized in a confined area. External forces and influence of inhomogeneities are taken into account as perturbations to exact solutions of the sine-Gordon equation. We have investigated two special cases of fluxon trapped by a microresistor and decay of a breather under dissipation. Also, we have carried out numerical simulations with dissipative sine-Gordon equation and made comparison with the McLaughlin-Scott theory. Significant distinction between the McLaughlin-Scott calculation for a breather decay and our numerical result indicates that the history dependence of the breather evolution can not be neglected even for small damping parameter.
 
 \end{abstract}
 \setcounter{page}{0}

\end{titlepage}

\newpage

\section{Introduction}
\indent\par 

Solitons or solitary waves are ones of the most interesting objects in nature. Observation of a solitary wave on water was first documented more than one and a half century ago. Besides, solitons occur naturally in many other substances like optical fibre~\cite{optic-fibre}, nonlinear lattices~\cite{lattices}, hot and cold plasma~\cite{Abdullaev} and are even mentioned responsible for Jupiter's red spots~\cite{Jupiter} and energy transfer in DNA~\cite{DNA}. Most intensively solitons have been studied in long Josephson contacts. The matter is that switching from a superconducting to resistive state of the Josephson junction is related to appearance and motion of solitons in these contacts which are known also as Josephson vortices or fluxons. Such solitons or fluxons are well described by the sine-Gordon equation.

In ideal case when the Josephson junction is infinitely long and narrow, Josephson solitons can be described analytically by well known exact solutions of the sine-Gordon equation.  
However, there is always dissipation associated with quasiparticle current through the Josephson junction and inhomogeneities associated with its width and thickness. Moreover, the real physical systems are always subjected to influence of external forces. All these factors may have significant impact on soliton behaviour. 

Although, the strictly one dimensional sine-Gordon equation is integrable~\cite{Ablowitz,Rajaraman}, the perturbations to this equation associated with the external forces and inhomogeneities spoil its integrability and the equation can not be solved exactly. Nevertheless, if their influence is small, the solution can be found perturbatively. The perturbation theory for solitons was described in details by Keener and  McLaughlin~\cite{Keener}. Later, in application to dynamics of vortices in Josephson contacts, the perturbation analysis of sine-Gordon equation has been developed by McLaughlin and Scott~\cite{McS}.

In many applications there appears a need in localized oscillatory solutions of sine-Gordon equation. For instance, when Josephson vortex is pinned by an inhomogeneity or there is a bound state of vortex with an antivortex known as breather. Breather may appear as a result of collision of a fluxon with an antifluxon or even in the process of measurements of switching current characteristics~\cite{Physica_C}. The role of breathers is ambiguous. Depending on our expectations, they can be parasitic excitations or, vice versa, a good substance for generation of THz waves. Recently we have proposed a device that may deliberatly generate and trap breathers~\cite{FCollider}. 

There have been many theoretical and numerical studies dedicated to continuous sine-Gordon breathers~\cite{decay-into-pair,ac-stabilization,boundaries,Quantized,others}. In particular, decay of a breather into fluxon and antifluxon induced by the external current was studied by many authors~\cite{decay-into-pair}. Moreover, it was shown~\cite{ac-stabilization} that a breather can be stabilized by ac drive even in presence of energy losses. Also, influence of the boundaries on a breather dynamics~\cite{boundaries} has been investigated and quantization of its energy spectrum~\cite{Quantized} has been predicted.

\begin{figure}[h!]
\begin{center}
\includegraphics[scale=0.9]{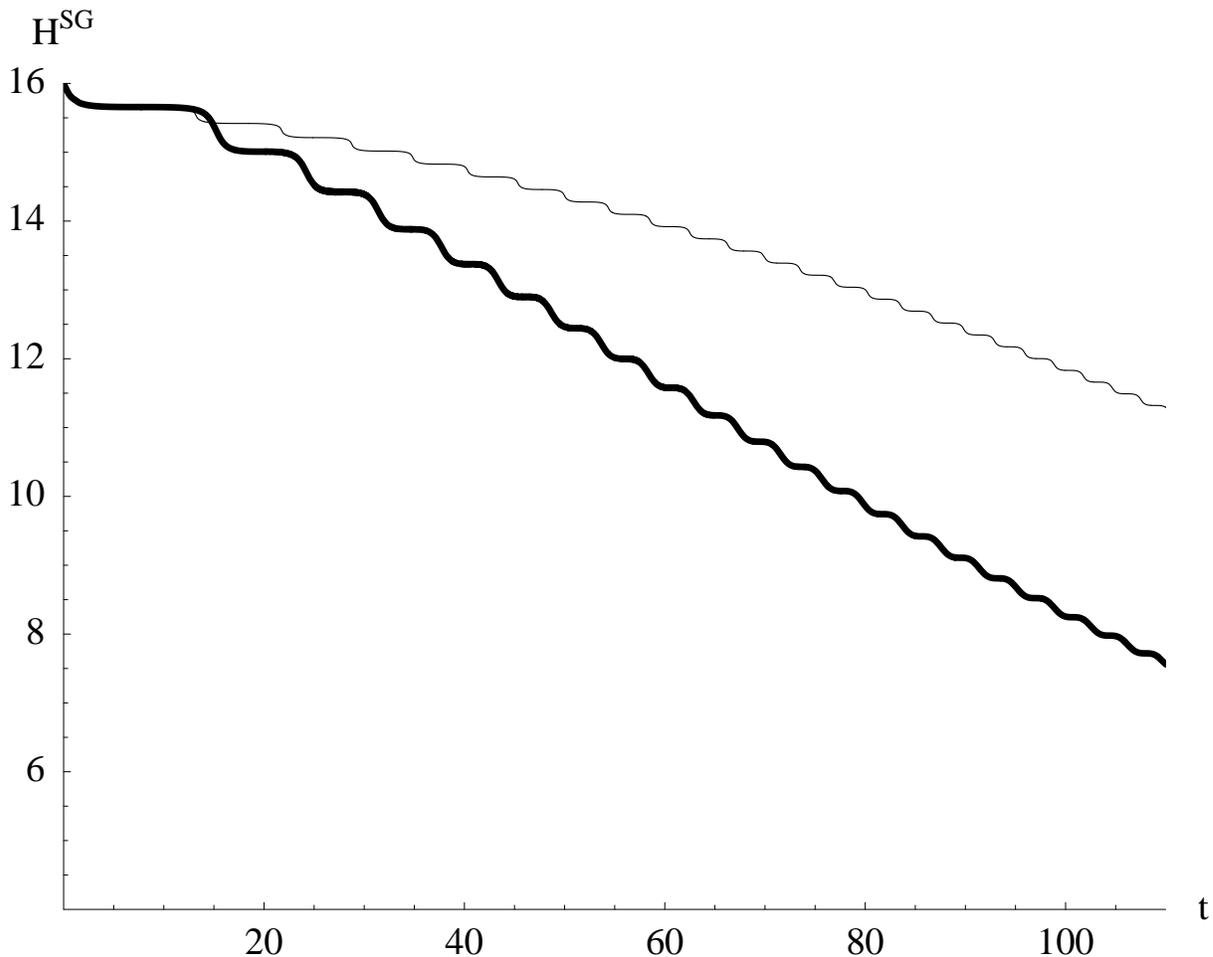}
\caption{\label{DecayMcL} Dissipative dynamics of a sine-Gordon breather: Dependence of the energy $H^{SG}$ of a breather on time $t$  calculated according to McLaughlin-Scott's formula (5.5) from the Ref.~\cite{McS}  is presented by thin line. Dependence of the energy $H^{SG}$ of a breather on time $t$   calculated by  direct numerical simulations of sine-Gordon equation with damping is shown by the thick solid line. The damping constant is $\alpha=0.01$.}
\end{center}
\end{figure}

Nevertheless, despite of numerous theoretical studies, the dynamics of a breather under dissipation has not been fully understood. McLaughlin-Scott theory gets overcomplicated when applied to non-trivial solutions such as breather, whereas its simplifications fail to predict the correct dynamics. We have performed numerical simulations of breather dynamics and found that there is a significant discrepancy with the McLaughlin-Scott calculation. In particular, it manifests itself in the dependence of the breather energy on time, Fig.\ref{DecayMcL}.  The thin line is the dependence following from McLaughlin-Scott calculation (formula (5.5) in Ref.~\cite{McS}), the solid line represents our numerical simulations. This discrepancy stimulated us to look into this problem once again and develop perturbation theory that is designed especially for localized solutions of sine-Gordon equation. We have found that at the construction of such theory it is very important to take into account the history dependence of the breather evolution. Also, we have carried out direct numerical simulations with dissipative  sine-Gordon equation. The numerical results appear in perfect agreement with our theory. 

\section{Perturbation theory for localized solutions}
\indent\par 
The reason why McLaughlin-Scott formula for breather decay (the formula (5.5) taken from the Ref.~\cite{McS}) fails to predict correctly the dissipative dynamics of a breather is the following. Breather is an oscillatory solution that is characterized by some "phase" that depends on history of the evolution. In their general formulation McLaughlin and Scott treat this difficulty introducing the history dependent term $\int_{t_0}^t\, u(t')\, dt'$ and allowing additional time-dependent modulation of the free parameters (such as initial positions of fluxons or phases of breathers) in the non-perturbed solution~\cite{Keener, McS}. The modulation of the free parameters is governed by additional differential equations. Obviously, this leads to additional complication because of the coupled differential equations for modulated parameters. Moreover, with such modulation the original solution no longer satisfies the non-perturbed sine-Gordon equation exactly and one needs also to modify the effective perturbation~\cite{McS}. Here we describe a method that does not involve the modulation of free parameters, but correctly deals with time-dependent dynamics due to appropriately chosen anzatz of non-perturbed sine-Gordon solution.

Consider solution of (1+1)-dimensional sine-Gordon equation
\begin{equation}
\ddot\varphi-\varphi_{xx}+\sin\varphi=0 
\label{non-pert-SG}
\end{equation}
in the form $ \varphi(g(u)\,x,\,g(u)\,u\,t,u) $ with $g(u)=1/\sqrt{1\pm u^2}$. Such parametrization is natural for sine-Gordon solutions such as solitons and their bound states. The sine-Gordon Hamiltonian is a functional of the field variable $\varphi$,
\begin{equation}
H^{SG}[\varphi]=\int_{-\infty}^\infty dx \left[\frac{\dot\varphi^2}{2}+\frac{\varphi_x^2}{2}+1-\cos{\varphi}\right]
\label{H_SG}
\end{equation}
Substitution of $\varphi=\varphi(g(u)\,x,\,g(u)\,u\,t,u)$ gives the effective energy as a function of a single parameter $u$,
$$H^{SG}_{eff}(u)=H^{SG}[\varphi(g(u)\,x,\,g(u)\,u\,t,u)]$$

The second argument of $\varphi(g(u)\,x,\,g(u)\,u\,t,u)$ that we call here a phase $T(t)=g(u)\,u\,t$ can be written in different ways, such as $T(t)=g(u)\,\int_0^t\, u\,dt'$ or $T(t)=\int_0^t\,g(u)\,u\,dt'$. Obviously, in case of $u$ independent on time these cases are  equivalent and the choice does not make any difference. However, this definition of the phase is very important when taking into account the influence of perturbations, as will be shown below.

In the presence of perturbations we assume that the dominant effect is to modulate the parameter $u=u(t)$. In other words, with appropriate choice of $u(t)$ we may satisfy the perturbed sine-Gordon equation
$$ \ddot\varphi-\varphi_{xx}+\sin\varphi=\epsilon f $$
by the function $\varphi=\varphi(g(u(t))\,x,T(t),u(t))$. Here,  we take the perturbation $\epsilon f$ in a general form
$$ \epsilon f=-\sum_i\mu_i\,\delta(x-x_i)\,\sin\varphi-\gamma-\alpha\,\dot\varphi$$
In contrast to the case of constant $u$, the choice of the non-perturbative solution is not unique anymore. Indeed, depending on the choice of the phase $T(t)$, we come out with different functions of $t$. We will show, that with the appropriate choice of the phase $T(t)$ we may correctly describe the time evolution of localized sine-Gordon solutions in presence of perturbations. We describe the dynamics by a single modulated parameter $u=u(t)$ without introducing additional modulation of the free parameters. This gives considerable simplification and improvement because the other free parameters such as initial location of solitons or initial phases of breathers remain fixed and do not result in auxiliary differential equations like those introduced in the Ref.~\cite{McS}.

Consider the anzatz 
\begin{equation}
\varphi(g(u(t))\,x,T(t),u(t))\quad \text{with}\quad T(t)=\int_0^t\,g(u(t'))\,u(t')\,dt'
\label{anzatz}
\end{equation}
where function $\varphi$ is an exact solution of non-perturbative sine-Gordon equation~\eqref{non-pert-SG}. In further consideration we omit highlighting the explicit dependence of the functions $u=u(t)$ and $T=T(t)$ for typographical convenience. Obviously, the drawback of the time modulation of $u$ affects the time derivative of $\varphi$,
$$ \dot\varphi=\frac{d}{dt}\varphi(g(u)\,x,T,u)=\varphi_1\,g_u\, \dot{u}\,x+\varphi_2\,g\,u+\varphi_3\,\dot u$$
Where $\varphi_1$, $\varphi_2$ and $\varphi_3$ are derivatives of $\varphi$ with respect to first, second and third argument correspondingly. As we consider localized solutions confined in some area $|x|<C$, the term $\varphi_1\,g'(u)\,\dot u\,x$ is of the order $O(\epsilon)$. The third term also can be neglected as it does not contain explicit linear terms in $x$ and $t$. Therefore,
\begin{equation}
\dot\varphi=\varphi_2\,g\,u+O(\epsilon)
\label{phi_t}
\end{equation}
that remains valid even in the limit of large times, $t\rightarrow\infty$. Obviously, another choice of $T(t)$ would spoil this equation with terms explicitly dependent on time $t$, e.g. for $T(t)=g(u(t))\,\int_0^t\,u(t')\,dt'$ we would have
$$
\dot\varphi(g(u)\,x, T, u)=\varphi_1\,g_u\, \dot{u}\,x+\varphi_2\,g\,u+\varphi_2\,g_u\,\dot u\,\int_0^t\,u(t')\,dt'+\varphi_3\,\dot u
$$
that contain a non-zero term $\int_0^t\, u(t')\,dt'$ proportional to $t$.
Thus, in this case the dynamics would not be correctly described on large time scales, $t\rightarrow\infty$. Mclaughlin and Scott overcome this problem introducing additional modulation of free parameters.

Substituting~\eqref{phi_t} to~\eqref{H_SG} we obtain the effective energy as a function of $u(t)$,
\begin{equation}
H^{SG}[\varphi(g(u(t))\,x,T(t),u(t))]\simeq H^{SG}_{eff}(u(t)) 
\label{H_SG_eff}
\end{equation}
that is valid for any values of $t$. It is important to note that this expression coincides exactly with the effective energy of non-perturbed solution~\eqref{H_SG} and depends on time inderectly only via $u(t)$.

In the presence of external forces, we may write the full Hamiltonian,
$$ H=H^{SG}+H^{P} $$
and take into account the dissipative perturbations affecting the energy dissipation rate~\cite{McS}
$$ \frac{dH}{dt}=-\int_{-\infty}^{\infty} \alpha \dot\varphi^2dx $$
The Hamiltonian $H^{P}$ serves to describe non-dissipative perturbations induced by external potential forces. This could be microshorts, microresistors or applied driving current.
$$
H^{P}=\int_{-\infty}^{\infty}\left(\sum_i\,\mu_i\,\delta(x-x_i)(1-\cos\varphi)+\gamma\varphi\right)dx
$$
Thus,
$$
\frac{dH^{SG}}{dt}=-\int_{-\infty}^{\infty}\left(\sum_i\,\mu_i\,\delta(x-x_i)\,\dot\varphi\,\sin\varphi+\gamma\dot\varphi+\alpha \dot\varphi^2\right)dx
$$
Substituting~\eqref{H_SG_eff}, we obtain the equation for parameter $u$,
\begin{equation}
\dot{u}\,\frac{dH^{SG}_{eff}}{du}=-\int_{-\infty}^{\infty}\left(\sum_i\,\mu_i\,\delta(x-x_i)\,\dot\varphi\,\sin\varphi+\gamma\dot\varphi+\alpha \dot\varphi^2\right)dx
\label{dotu}
\end{equation}
where $\varphi=\varphi(g(u)\,x,T)$ and $\dot\varphi\simeq\varphi_T\,g(u)\,u$ should be substituted. This equation is coupled to the equation for~$T$,
\begin{equation}
\dot T=g(u)\,u
\label{dotT}
\end{equation}
In some cases it can be convenient to rewrite this system of differential equations for independent variable~$T$,
\begin{equation}
\begin{cases}
\frac{du}{dT}=-\left(\frac{dH^{SG}_{eff}}{du}\,g(u)\, u\right)^{-1} \int_{-\infty}^{\infty}\left(\sum_i\,\mu_i\,\delta(x-x_i)\,\dot\varphi\,\sin\varphi+\gamma\dot\varphi+\alpha \dot\varphi^2\right)dx
\\
\frac{dt}{dT}=\left(g(u)\, u\right)^{-1}
\end{cases}
\label{systemT}
\end{equation}
The dynamics is described by $u(T(t))$, where $T(t)$ is a inverse function of $t(T)$.

\section{Pinning by a microresistor}
\indent\par 
Let us consider the following ansatz for a single soliton solution 
$$ 
\varphi(g(u))\,x,T)=4\arctan\exp\left(g(u) x - T\right),
\quad
g(u)=1/\sqrt{1-u^2}
$$
subjected to the attractive potential of a microresistor
$$ 
\epsilon f=-\mu\,\delta(x)\,\sin\varphi,\quad \mu<0
$$
The energy of a soliton is equal to
$$ H^{SG}_{eff}(u)\simeq \frac{8}{\sqrt{1-u^2}} $$
From~\eqref{dotu} and~\eqref{dotT} we obtain the next system of coupled differential equations,
\begin{equation}
\begin{cases}
\dot{u}=\frac12 \mu\, (1-u^2)\, \sech^2(T(t))\,\tanh(T(t))
\\
\dot T=\frac{u}{\sqrt{1-u^2}}
\end{cases}
\label{microresistor}
\end{equation}
We have found that after some simplifications, the McLaughlin-Scott's formula (4.3) from Ref.\cite{McS} can be reduced to the exactly the same system of differential equations. 
Although,  both approaches lead to exactly the same result, McLaughlin-Scott's formulation is, obviously, more cumbersome.

\begin{figure}[h!]
\begin{center}
\includegraphics[scale=1.4]{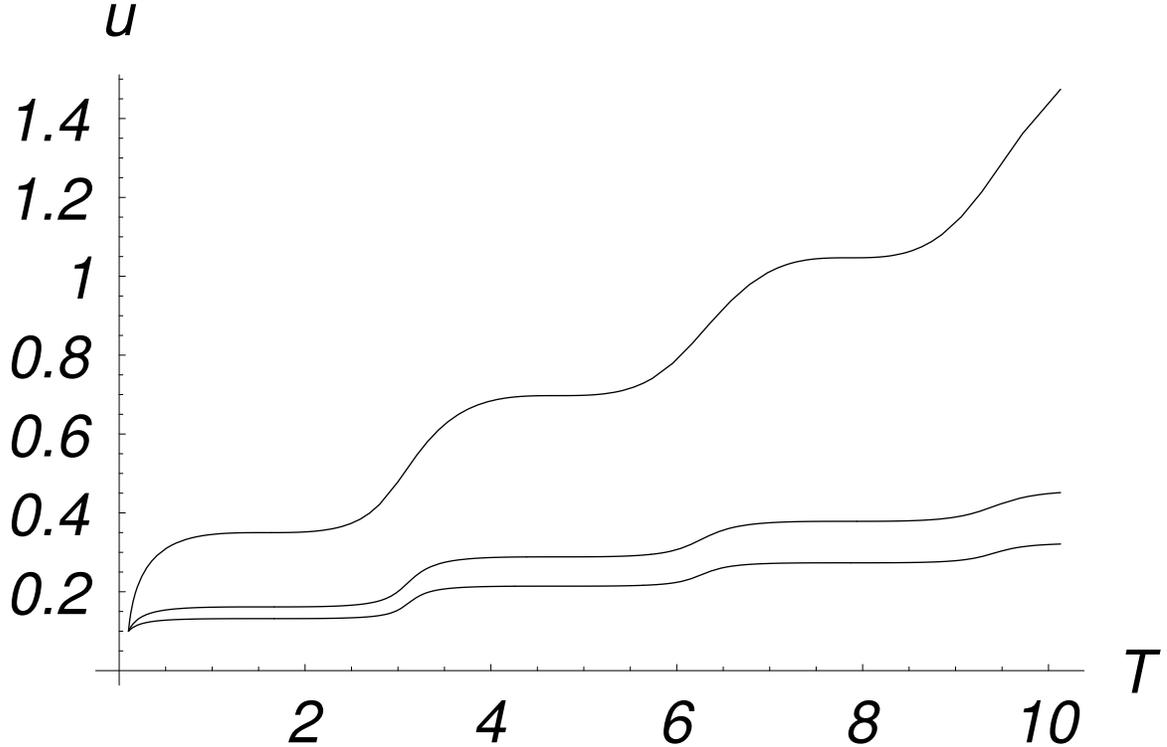}
\caption{\label{u[T]} 
The dependence of the speed $u(T)$  on the phase $T$  calculated using perturbation theory for localized sine-Gordon solutions at different damping rates. The top curve corresponds
to the damping constant is $\alpha=0.05$, the middle curve - $\alpha=0.01$ and the lower curve $\alpha=0.005$ . }
\end{center}
\end{figure}

\begin{figure}[h!]
\begin{center}
\includegraphics[scale=1.6]{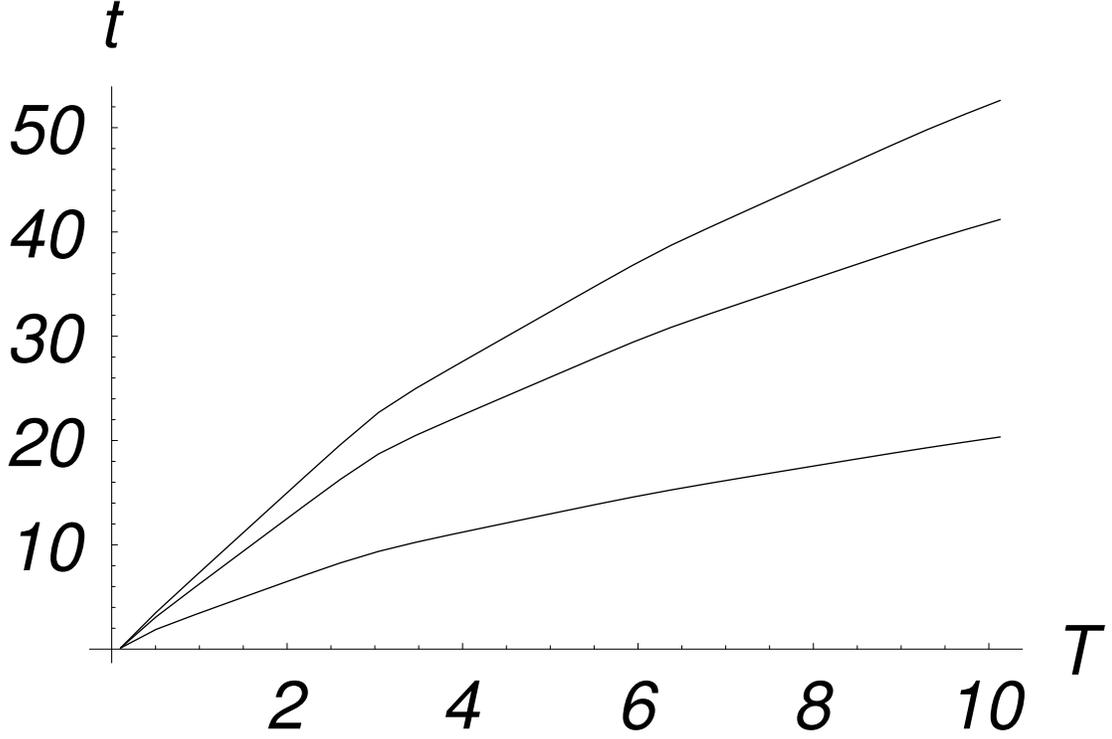}
\caption{\label{t[T]} 
The dependence of the time $t$  on the phase $T$  calculated using perturbation theory for localized sine-Gordon solutions at different damping rates. The top curve corresponds
to the damping constant is $\alpha=0.005$, the middle curve - $\alpha=0.01$ and the lower curve corresponds to the highest damping $\alpha=0.05$ .}
\end{center}
\end{figure}

\section{Decay of a breather}
\indent\par 
Consider a breather solution
$$ \varphi(g(u)\,x,T,u)=4\arctan\left(\frac{\sin T}{u \cosh \left(g(u) x\right) }\right) $$
with $g(u)=1/\sqrt{1+u^2}$. 
As a perturbation we consider the dissipative term
$$ \epsilon f=-\alpha\,\dot\varphi$$
The effective energy is
$$ H^{SG}_{eff}(u(t))\simeq \frac{16}{\sqrt{1+u(t)^2}} $$

\begin{figure}[h!]
\begin{center}
\includegraphics[scale=0.9]{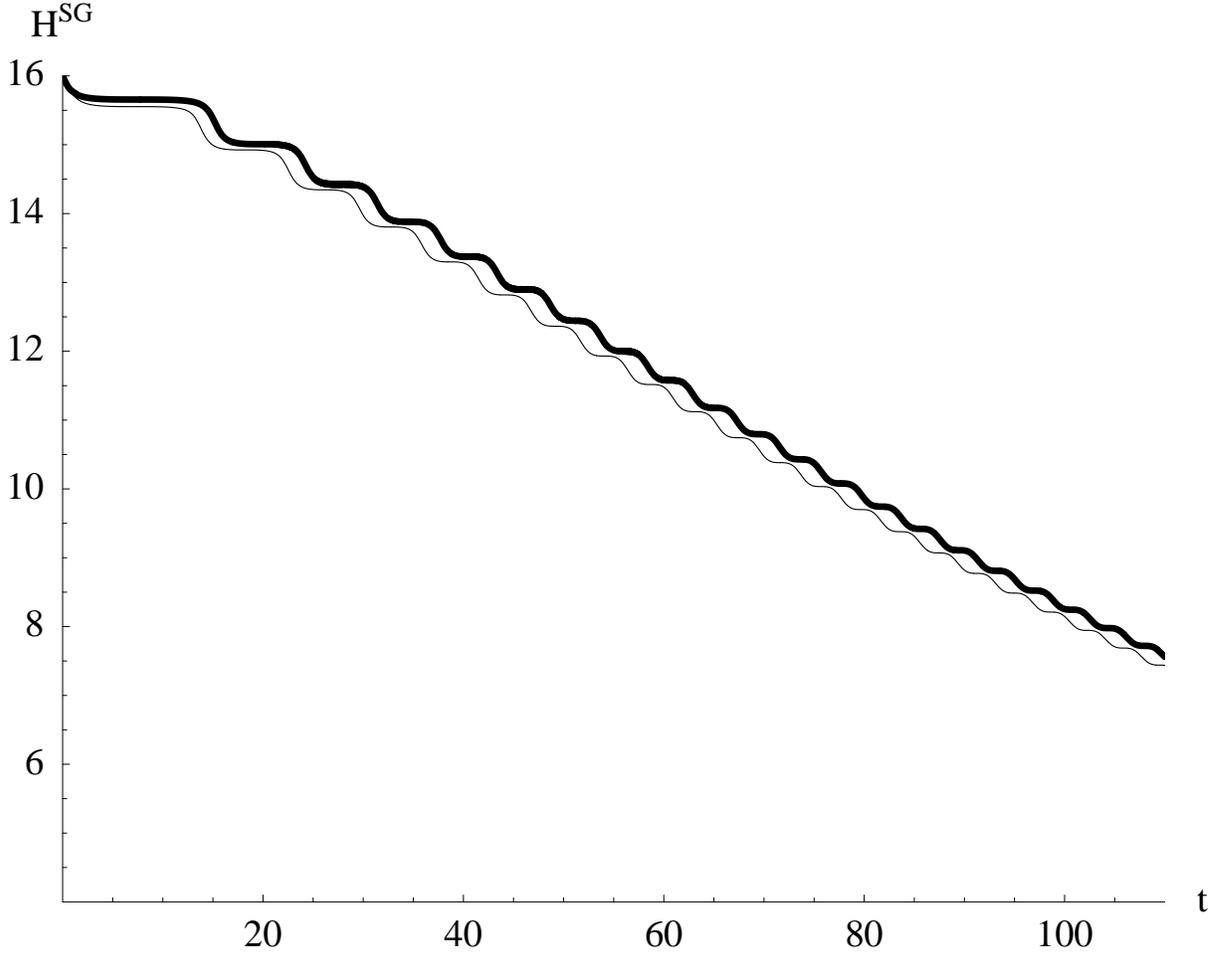}
\caption{\label{DecayMy} 
Dissipative dynamics of a sine-Gordon breather: Dependence of the energy $H^{SG}$ of a breather on time $t$  calculated using perturbation theory for localized sine-Gordon solutions (thin line). Dependence of the energy $H^{SG}$ of a breather on time $t$ calculated by  direct numerical simulations of sine-Gordon equation with damping is shown by the thick solid line. The damping constant is $\alpha=0.01$. }
\end{center}
\end{figure}

From~\eqref{systemT} we obtain the next system of coupled differential equations
\begin{equation}
\begin{cases}
\frac{du}{dT}= \alpha\, \frac{(1+u^2)^{3/2}\,\cos^2 T}{\sin^2 T+u^2}\left[1+\frac{u^2\,\arctanh \left(\frac{\sin T}{\sqrt{\sin^2 T+u^2}}\right) }{\sin T \, \sqrt{\sin^2 T+u^2}}\right]
\\
\frac{dt}{dT}=\frac{\sqrt{1+u^2}}{u}
\end{cases}
\label{breather-decay}
\end{equation}
where $u=u(T(t))$. This is a new result that may not be obtained from the McLaughlin-Scott theory by straightforward manipulation. The system can be solved numerically.  For an illustration we present the solution of these equations on Figs~\ref{u[T]} and \ref{t[T]}. There the initial conditions are taken as $u(0.1)=0.1$ and $t(0.1)=0.1$. One may notice the steplike character of the dependence $u(T)$, see Fig.\ref{u[T]}. The size of the steps increases with dampling indicating importance of the introduction of the phase $T(t)$. This phase has also nontrivial dependence on time $t$. Its inverse function $t(T)$ is presented in Fig.\ref{t[T]}. One may notice that at some values of $T$ there is a fast change of the slope. Obviously this is related to the steplike character of the dependence $u(T)$.
The dissipative dynamics of a breather is also well reflected by time dependence of its energy,~Fig.\ref{DecayMy}. The results are in perfect agreement with our numerical simulations using the complete sine-Gordon equation with dissipative term.

\section{Conclusion}
\indent\par 

In summary, we have found that our perturbation theory describes well the dynamics of localized excitations subjected to influence of external forces such as various inhomogeneities and damping associated with quasiparticle current. In particular, we have described a fluxon trapped in a potential well which could be related to a microresistor in the Josephson junction.  Here the equations  derived with the use of our method coincides identically  with equations derived by McLaughlin-Scott~\cite{McS}. However the derivation of these equations obtained by our method is significantly simpler. Second,  we have described the decay of the breather under dissipation. In this case, the equations are different from the McLaughlin-Scott's~\cite{McS}. According to our calculation, the breather is decaying significantly faster. In order to resolve this difference we have performed numerical simulations with dissipative sine-Gordon equation. The results of these numerical simulations are in perfect agreement with our theoretical results. The comparison of our perturbation theory with the McLaughlin-Scott's calculation~\cite{McS} indicates that the history dependence of the breather evolution has a strong influence on its dynamics even at low damping.

To conclude, we have developed the perturbation theory which perfectly describes 
 the localized in space solutions of sine-Gordon equation. Such study can be important for new devices such as fluxon collider~\cite{FCollider} or other devices based on newly discovered flux cloning effect~\cite{Cloning} where the dissipative dynamics and the breather excitations may play a key role. The theory may allow generalization to higher dimensions. This can be of use to study localized pulsating solutions of sine-Gordon in two spacial dimensions~\cite{Zakrzewski}.

\end{document}